\input epsf
\documentstyle[11pt]{article}

\def\beq{\begin{equation}}
\def\eeq{\end{equation}}

\def\half{\frac{1}{2}}
\def\third{\frac{1}{3}}

\def\ph3{$\phi^3$}
\def\gst{\gamma_{str}}

\def\mgap{\qquad \qquad}

\def\gone{{{\cal G}_r^I}}
\def\zone{{\cal Z}^{I}}
\def\xcone{{x_c^{I}}}

\def\gtwo{{{\cal G}_r^{I\!I}}}
\def\ztwo{{\cal Z}^{I\!I}}
\def\xctwo{{x_c^{I\!I}}}

\def\nmean{{\langle n \rangle}}

\def\zgau{{{\cal Z}_g^{I}}}
\def\zgautwo{{{\cal Z}_g^{I\!I}}}
\def\span{{{\cal T}_G}}

\def\ltsim{\lower3pt\hbox{$\, \buildrel < \over \sim \, $}}

\begin{document}

\vbox{\smash{\vbox{
\begin{flushright}
 \large NBI-HE-96-37 \\
 \large July 16, 1996
\end{flushright}}}
\title{
The Branching of Graphs in 2-d Quantum Gravity} 
\author{M. G. HARRIS \\ \\
  \small Niels Bohr Institute, Blegdamsvej 17, \\
  \small DK-2100 Copenhagen, Denmark. \\ 
  \small E-mail address: Martin.Harris@nbi.dk}
\date{}
\maketitle}
\begin{abstract}
The branching ratio is calculated for three different models of 2d
gravity, using dynamical planar $\phi^3$ graphs. These models are pure
gravity, the $D\!=\!-2$ Gaussian model coupled to gravity and the
single spin Ising model coupled to gravity. The ratio gives a measure
of how branched the graphs dominating the partition function
are. Hence it can be used to estimate the location of the branched
polymer phase for the multiple Ising model coupled to 2d gravity.
\end{abstract}


\section{Introduction}

Many interesting results in the area of 2d quantum gravity have been
obtained by studying models of matter fields coupled to dynamical
triangulations. However, all the exactly solved models have had
central charges, $C$, of less than or equal to one (if one ignores
models for which $C= \infty$~\cite{Wexler,ADJ}).
Using conformal field theory it is possible to make predictions for
the critical exponents in models of matter coupled to 2d
gravity~\cite{KPZ,DDK}, the so-called ``KPZ formulae'', but again
these formulae are only valid for $C \le 1$. The nature of this
``$C=1$ barrier'' remains something of a mystery. Many models of
matter coupled to gravity have a tendency to degenerate into branched
polymer configurations for large enough $C$. It is thus tempting to
hypothesise that a branched polymer phase begins at $C=1$, causing
the break down of the KPZ formulae here. Models in which multiple
Ising spins are coupled to dynamical triangulations (or their dual
\ph3 graphs) provide a convenient way of testing this hypothesis; such
a model with $p$ independent Ising spins on each vertex of the graph, has
a central charge of $C=p/2$ associated with it.
However, Monte Carlo simulations for
values of $p$ up to $p=16$~\cite{MC} have failed to show any
convincing evidence of the existence of such a branched polymer (``BP'')
phase. Since there are reasons to believe that such a phase
exists~\cite{Wexler,paper2}, there are two possibilities to consider:
firstly, that there exists an intermediate region between $C=1$ and
the start of the branched polymer phase, or secondly, that the BP
phase does indeed start at $C=1$ and that the MC simulations have for
some reason failed to detect it. This paper aims to throw some light
on the question by examining the branchedness of the graphs for
various exactly solvable models.

In a recent paper~\cite{paper4} we studied correlation functions in a
model of multiple Ising spins on a dynamical planar \ph3 graph. 
In the paper, a function $B$ was defined which played a key r{\^o}le
in the calculation of the critical exponents;
$B$ gives a measure of how branched the \ph3 graphs
contributing to the partition function are and hereafter it will be
called the ``branching ratio''.
The branching ratio lies in the range $0$ to $1$,
and throughout the branched polymer 
phase $B$ equals one, when evaluated at the critical value
of the cosmological constant. Thus the branching ratio gives a measure
of how close to the BP phase we are, and by evaluating it for various
models we will gain some indication of the extent of the branched
polymer phase for the multiple Ising model coupled to gravity.

In this paper we calculate the branching ratio, which is defined in
the next section, for three different
models (all of which are defined in terms of planar \ph3 graphs): 
pure gravity (section~\ref{sec:puregrav}), the $D \!= \! -2$ Gaussian
model coupled to gravity (section~\ref{sec:gaussian}) and the single
spin Ising model coupled to gravity (section~\ref{sec:ising}). As
might be expected the branching ratio increases with increasing
central charge. In section~\ref{sec:concl} we conclude by using these
results to make a prediction of the location of the branched polymer
phase for the multiple Ising model coupled to gravity; it is estimated
that there exists a BP phase for central charges above some value
$\overline{C}$,
where $\overline{C}$ lies in the range $1 \le \overline{C} \ltsim 2.75$.

\section{The branching ratio}
\label{sec:branch}

\begin{figure}[bth]
\caption[l]{Measuring distances from the root blob}
\label{fig:measure}
\begin{picture}(100,55)(0,0)
\centerline{\epsfbox{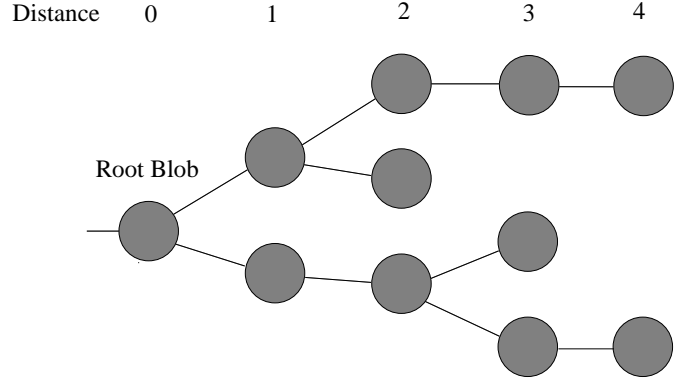}}
\end{picture}
\end{figure}

In our previous paper~\cite{paper4}
we used a simplified definition of distance rather than the usual
geodesic distance. The geodesic distance between two vertices $A$ and $B$
of a \ph3 graph is defined to be the shortest path along links, from
$A$ to $B$, counting one unit of distance for each link
traversed. However, in our definition of distance, only links
separating two one-particle irreducible (``1PI'') 
subgraphs are counted; figure~\ref{fig:measure}
gives an example, each shaded circle represents a 1PI subgraph,
hereafter referred to as a ``blob''.

Using this definition of distance it was possible to show that the
number of blobs at a distance $r$ from the root, denoted $\nmean_r$, was given
by $\nmean_r=B^r$, where $B$ is a function of the various parameters
in the model. 

For the case of a single Ising spin coupled to each
vertex of a graph (using the set, $\gone$, of all rooted planar \ph3
graphs), we derived the following formula for
$B$~\cite{paper4},
\beq
\label{eqn:fullB}
B(x,\beta)= \frac{x T}{1-2xT} \left[ 1 -3xT + \frac{(1-xT)}{\ztwo} \left( 
3 x' \left(\frac{\partial \ztwo}{\partial x'}\right)_{\! \! \! \beta'} + 
2 t' ( t' -1 ) \left(
\frac{\partial \ztwo}{\partial t'} \right)_{\! \! \! x'} \right) \right],
\eeq
where $T\equiv \zone(x,\beta)$. The grand canonical partition function
for the single spin Ising model on the set of graphs $\gone$
is defined by
\beq
\label{eqn:general}
\zone(x,\beta) = \sum_{G \in \gone} x^N \frac{1}{2^{N}}
\sum_{\{S\}} \prod_{<ij>} \left[ 1 + t
S_i S_j \right] ,
\eeq
where $S_i$ is the  spin on vertex $i$ and
$t=\tanh \beta$. This will be referred to as model~I.
The coupling constant $\beta$ is the inverse
temperature of the model and each vertex in the graph is weighted with
a factor of $x$. The first summation is over graphs $G$ in the set
$\gone$, where $N$ is the number of vertices in $G$,
the second summation is over all spin configurations
and the product
is over nearest neighbour pairs. Note that there is no spin at the end
of the root and hence the root link is not included in this product.
The partition function
$\ztwo \equiv \ztwo(x',\beta')$ is defined as in (\ref{eqn:general}),
but using the set of planar rooted 1PI \ph3 graphs, $\gtwo$; 
this will be called model~II. The renormalized
coupling constants $t'=\tanh \beta'$ and $x'$ are given by
\beq
\label{eqn:fullmap}
x' = x(1-2xT)^{-\frac{3}{2}} , \mgap
t' = t \frac{(1-2xT)}{(1-2xT t)} 
\eeq
and the following equation which relates $\zone$ and $\ztwo$ was
derived in our paper~\cite{paper4},
\beq
\label{eqn:fullf}
T=\sqrt{1-2xT} \ \ztwo(x',\beta') + x T^2 .
\eeq

\section{Pure gravity}
\label{sec:puregrav}
In the case of pure gravity (that is, equation (\ref{eqn:general}) with
 $\beta=0$) equation (\ref{eqn:fullB}) reduces to
\beq
\label{eqn:PGB}
B(x)= \frac{x T}{1-2xT} \left[ 1 -3xT + 3(1-xT)\frac{x'}{\ztwo} 
 \frac{\partial \ztwo}{\partial x'}  \right].
\eeq
We are interested in evaluating this at the critical value of $x$ for
model~I,
which is denoted by $\xcone$. 
From reference~\cite{BIPZ}, $\xcone= 1/(2.3^{\frac{3}{4}})$
and $T(\xcone) = 3^{\frac{3}{4}}\left( 1 - \sqrt{3}/2 \right)$.
The corresponding value of $x'$ is $\xctwo= \sqrt{\frac{2}{27}}$ and
in reference~\cite{paper4} it was shown that
\beq
\frac{x'}{\ztwo} \left. \frac{\partial \ztwo}{\partial x'}
\right\vert_{x'=\xctwo} = \frac{5}{3} .
\eeq

Thus writing $B_c$ for $B(\xcone)$, we have 
that the branching ratio at the critical value of $x$ is
\beq
B_c =1- \frac{1}{\sqrt{3}} \approx 0.4226
\eeq
for pure gravity, with rooted planar
\ph3 graphs; this model, of course, corresponds to a central charge of zero.

\section{Gaussian model}
\label{sec:gaussian}
In this section we will derive $B_c$ for the Gaussian model in
$D\!=\!-2$ dimensions, which has a central charge of $-2$. 
For this choice of $D$ the partition function
has an especially simple form~\cite{BKKM},
\beq
\zgau(x) = \sum_{G \in \gone} \sum_{T \in \span} x^N .
\eeq
Again the set $\gone$ of all rooted planar \ph3 graphs is being used
and in addition there is a summation over the trees $T$ in the set of all
trees spanning $G$; this set is denoted by $\span$.

The partition function has an integral
formulation~\cite{KKM},
\beq
\zgau(x) = \frac{1}{x \pi} \int_{-1}^{1} dt \sqrt{1-t^2} \left( 1
- \sqrt{1-8xt} \right) ,
\eeq
which has a critical value of $x$ given by $\xcone=\frac{1}{8}$. At this
critical value,
\beq
\zgau(\xcone) = 4 \left( 1 - \frac{32 \sqrt{2}}{15 \pi} \right),
\eeq
\beq
\left. \frac{d \zgau}{dx} \right\vert_{\xcone} = 32 \left( \frac{12
\sqrt{2}}{5 \pi} -1 \right)
\eeq
and $\frac{d^2 \zgau}{dx^2}$ diverges.

\begin{figure}[bth]
\caption[l]{$T=f(T)$}
\label{fig:tft}
\begin{picture}(100,25)(0,0)
\centerline{\epsfbox{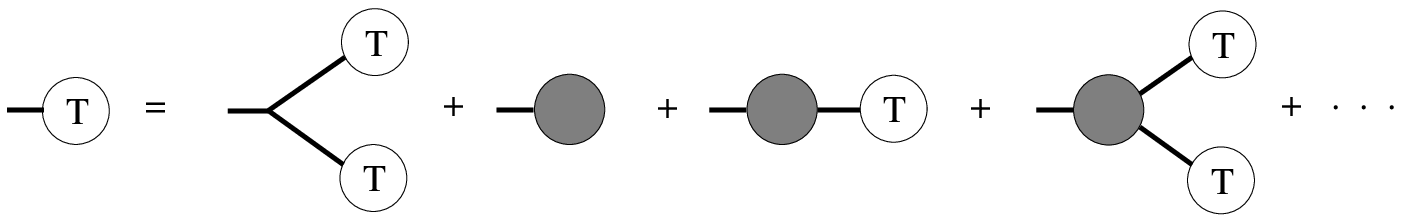}}
\end{picture}
\end{figure}
  
\begin{figure}[bth]
\caption[l]{Renormalization of links: (a) Black, (b) White}
\label{fig:blwh}
\begin{picture}(100,55)(0,0)
\centerline{\epsfbox{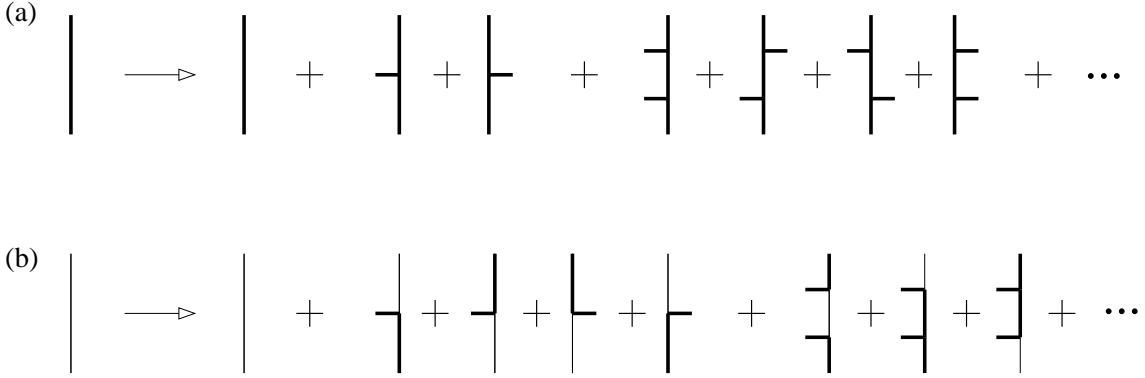}}
\end{picture}
\end{figure}

A formula for $B(x)$ in this model can be obtained by using a 
derivation similar to that for the multiple Ising model~\cite{paper4}.
Define a function, $f(y)$, which takes a rooted 1PI blob (both graphs
and spanning trees are being summed over in the blob) and glues an
arbitrary number of legs, each weighted with a factor $y$, on to the
blob. Again we define $T \equiv \zgau(x)$ to save writing and have
that $T=f(T)$, which is illustrated in figure~\ref{fig:tft}. To derive
a formula for $f(y)$ we need to consider the effect of adding an
arbitrary number of legs to a link. For a given graph $G$ and spanning
tree $T$ there are two different types of link, those that form
part of the spanning tree, which will be referred to as ``black''
links (drawn as thick lines), and those which are not part of the
spanning tree (these are called ``white'' and drawn as thin lines).
Adding legs to a black link changes the contribution it makes to the
partition function (fig~\ref{fig:blwh}a):
\beq
1 \to 1 + 2xy + (2xy)^2 + \cdots = \frac{1}{1-2xy}.
\eeq
For each leg added there is a weight of $y$, a weight of $x$ for the
extra vertex and a factor of $2$, since the leg can be hung from the
link in one of two directions; this accounts for the factors of $2xy$.

Similarly for white links there is the change (fig~\ref{fig:blwh}b):
\beq
1 \to 1 + 2(2xy) + 3(2xy)^2 +4 (2xy)^3 + \cdots = \frac{1}{(1-2xy)^2} .
\eeq
In this case one must also take into account the different ways of
connecting the new legs to the existing spanning tree.

The partition function for rooted 1PI blobs is defined by 
\beq
\zgautwo(x) = \sum_{G \in \gtwo} \sum_{T \in \span} x^N.
\eeq
Adding legs to this changes the partition function for the blob to
\beq
\sum_{G \in \gtwo} \sum_{T \in \span} x^N (1-2xy)^{-(L_B+2L_W)},
\eeq
where the number of black links is $L_B=N-1$ and the number of white
links is $L_W=\half(N+1)$. Thus $L_B+2L_W=2N$ and hence
\beq
f(y)= xy^2+ \zgautwo \left(x(1-2xy)^{-2} \right).
\eeq
It is convenient to define $x'= x(1-2xT)^{-2}$. Note that at
$x=\xcone$, we have $x'=\xctwo \equiv \left(\frac{15 \pi}{128}
\right)^2 $, which agrees with the result in reference~\cite{Dav85}.
As before $B$ is given by
\beq
B= \left. \frac{\partial f}{\partial y} \right\vert_{y=T},
\eeq
so that, with $\zgautwo \equiv \zgautwo(x')$,
\beq
\label{eqn:Bgau}
B(x)= 2 x \left[ T+ \frac{ d \zgautwo}{d x'} \frac{2 x'}{1-2xT} \right].
\eeq
Differentiating the equation $T=f(T)$ with respect to $x'$ yields
an expression for
$\frac{d\zgautwo}{dx'}$ and then evaluating (\ref{eqn:Bgau}) at
$x=\xcone$ gives
\beq
B_c= 1- \frac{3 \pi}{8 \sqrt{2}} \approx 0.1670,
\eeq
for the $D\!=\!-2$ Gaussian model.

\section{Ising model}
\label{sec:ising}

Finally, we will calculate the branching ratio for the Ising model on 
planar \ph3 graphs with a single spin per vertex; the matter in 
this model has a
central charge of $\half$. To use equation (\ref{eqn:fullB}) we need the
partition functions for both model~I and model~II, which we obtain by
solving the following two-matrix model,
\begin{eqnarray}
{\cal Z} = e^{M^2 F(g,c,\lambda)} & = & \int {\cal D} A \ {\cal D} B \ \exp
\mathop{\rm Tr} \Biggl[  - \half (A^2+B^2) + c AB  \nonumber \\
& + & \third \frac{g}{\sqrt{M}} \left(A^3 +B^3\right) 
 +  \lambda \sqrt{M}
\left(A+B\right) \Biggr] ,
\end{eqnarray}
where $A$ and $B$ are $M \times M$ Hermitian matrices and we wish to
take the planar limit in which $M \to \infty$. Each vertex is weighted
with $g$ and the coupling constant $c$ equals $e^{-2\beta}$. In the
case $\lambda=0$, $F$ will give the model~I partition function up to
various factors; this has been solved in references~\cite{BouKaz1,BurJur}.
Alternatively by choosing $\lambda$ so that graphs with tadpoles are
cancelled, we gain the partition function for model~II.

This matrix model can be solved using orthogonal
polynomials~\cite{Mehta}. In the planar limit and dropping the
constant term, we have
$$
F(g,c,\lambda) = \half \ln \left(\frac{1}{g^2} z(u=1) \right) +
\int_0^1 \left(\half u^2 -u \right) \frac{1}{z}\frac{dz}{du} du
+\frac{3}{4} + \half \ln \left(1-c^2 \right) $$
\beq
- \third \left(
\frac{1}{2g} \right)^2 \left(1+\rho_0 \right) \Bigl[ \left(1-c \right)
\left(1+\rho_0\right) -
8 \lambda g \Bigr],
\eeq
where
\beq
 u g^2 = z (\rho^2 -c^2) +2cz^2 ,
\eeq
\beq
z=\frac{1}{8 \rho} \Bigl[\left(\rho+1\right) \left(\rho-1+2c\right)
 +4 \lambda g \Bigr]
\eeq
and $\rho_0$ is $\rho$ evaluated at $u=0$ (i.e.\ at $z=0$),
\beq
\rho_0= -c- \sqrt{(1-c)^2 - 4 \lambda g} .
\eeq
After some considerable algebra we obtain,
\beq
g \left(\frac{\partial F}{\partial \lambda}\right)_{g,c} = 1+ \rho_1 -
\frac{2}{g^2} z_1^2 \rho_1 (\rho_1 -c) ,
\eeq
where $\rho_1=\rho(u=1)$ and $z_1=z(u=1)$.
Note that $F$ generates unrooted (i.e.\ vacuum) graphs and that by
differentiating with respect to $\lambda$ we generate rooted graphs
instead. 

Now, the model~I partition function that we want to calculate is related to
that generated from the matrix model by 
\beq
xT \equiv x \zone(x,\beta) = \half \left( \frac{g}{1-c} \right)
\left. \frac{\partial F}{\partial \lambda} \right\vert_{\lambda=0}.
\eeq
Thus dropping the subscripts,
\beq
\label{eqn:xT}
xT = \half \frac{(1+ \rho)}{(1-c)} \left[ 1 - \frac{\rho (\rho -c) (\rho-
\alpha)}{(4 \rho^3 +c \rho^2 - 2 c^2 \rho -c \alpha)}\right],
\eeq
where $c=e^{-2\beta}$, $\alpha=1-2c$,
\beq
z= \frac{1}{8 \rho} (\rho+1) (\rho- \alpha),
\eeq
\beq
g^2 = z(\rho^2 -c^2) + 2c z^2 = 2 (1-c)^3 x^2.
\eeq
At $\xcone$, which is given by $\left(\frac{\partial g}{\partial
\rho}\right)_c =0$,
\begin{eqnarray}
\rho = - \sqrt{\third(1-2c)} \mgap &({\rm for} \ \beta > \beta^*) ,\\
2 \rho^3 +3c \rho^2 +c (1-2c) =0 \mgap &({\rm for} \ \beta < \beta^*) .
\end{eqnarray}
There is a third order phase transition in model~I, at
\beq
c^*= \frac{1}{27}\left(2\sqrt{7} -1 \right),
\eeq
between a magnetized and an unmagnetized phase.

\begin{figure}[bth]
\caption[l]{Graph of the branching ratio, $B_c(\beta)$, against
 $c=e^{-2\beta}$ for the
single spin Ising model coupled to 2d gravity.}
\label{fig:Bcvsc}
\begin{picture}(100,100)(0,0)
\centerline{\epsfbox{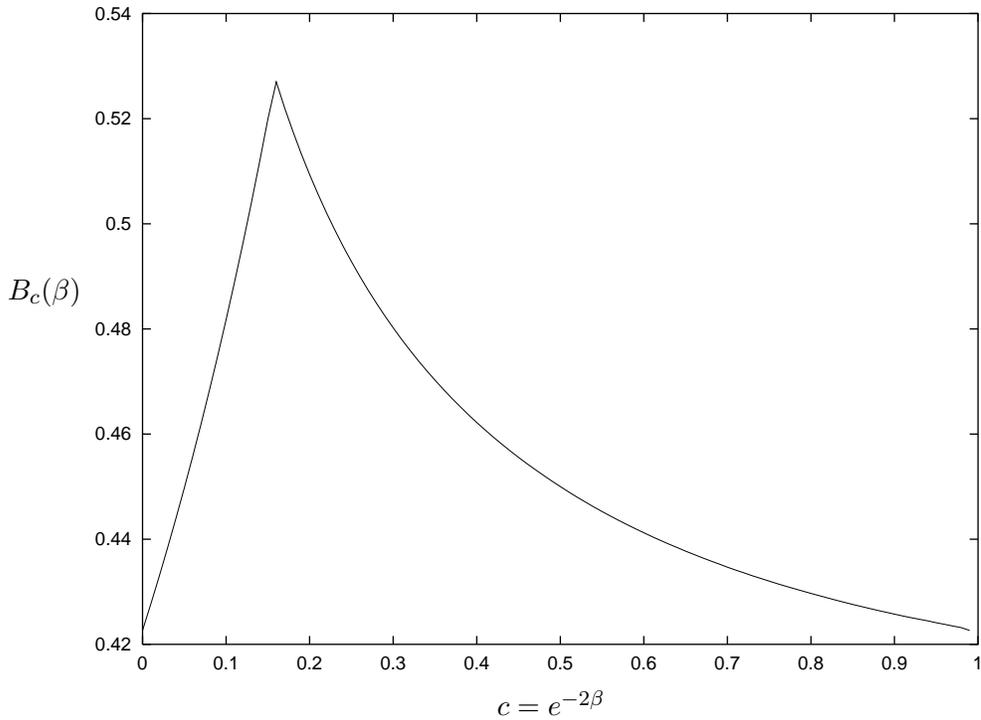}}
\end{picture}

\begin{picture}(10,10)(-5,-60)
$B_c(\beta)$
\end{picture}

\begin{picture}(10,10)(-70,-15)
$c=e^{-2\beta}$
\end{picture}
\end{figure}

For model~II we need to choose $\lambda=\overline{\lambda}$ such that
$\left.\frac{\partial F}{\partial
\lambda}\right\vert_{\overline{\lambda}}$ vanishes. This gives us,
\beq
\label{eqn:gg}
g^2 =  \frac{\rho (1+\rho) (\rho+c)^2 (\rho-c)^3}{2 (\rho^2-2 \rho
c - c )^2}
\eeq
and
\beq
\label{eqn:lg}
\overline{\lambda} g = \frac{1}{4} \frac{(1+\rho)}{(\rho^2-2 \rho c -
c)} \left(3 \rho^3 + \rho^2 - \rho c - c (1-2c) \right).
\eeq
At the critical value of $x$, denoted $\xctwo$,
\begin{eqnarray}
3\rho^2 +2 \rho +c = 0 \mgap & ({\rm for} \ \beta>\beta^*), \\
\rho^3 -3 c \rho^2 - 3 c \rho +c^2 =0 \mgap & ({\rm for} \ \beta<\beta^*).
\end{eqnarray}
There is a third order phase transition at $c^*=\frac{5}{27}$, and the
model~II partition function is
\beq
\label{eqn:ztwo}
x \ztwo(x,\beta) = - \frac{\overline{\lambda} g}{(1-c)^2} .
\eeq
Defining
\beq
E(x,\beta) \equiv \frac{1}{\ztwo(x,\beta)} \left[ 3 x \left(\frac{\partial
\ztwo}{\partial x}\right)_{\!\!\!\beta} + 2 t (t-1) \left( 
\frac{\partial \ztwo}{\partial t}\right)_{\!\!\!x}
\right] ,
\eeq
we have from equations (\ref{eqn:gg}), (\ref{eqn:lg}) and (\ref{eqn:ztwo})
that
\beq
E(x,\beta) = 4c-3 + \frac{4(1-c)(3 \rho^3 +c^2)}{(3 \rho^3 + \rho^2 -
\rho c - c(1-2c))} .
\eeq
Then
\beq
B(x,\beta) = \frac{xT}{1-2xT} \Bigl[ 1 -3xT + (1-xT) E(x', \beta') \Bigr],
\eeq
where $E$ is given by the previous equation, $x'$ and $\beta'$ by
(\ref{eqn:fullmap}) and $xT$ is given by (\ref{eqn:xT}).
Evaluating this at the critical value of $x$ gives $B_c(\beta) \equiv
B(\xcone(\beta),\beta)$; figure~\ref{fig:Bcvsc} gives a plot of $B_c$
against $c=e^{-2\beta}$. The branching ratio is a maximum at the phase
transition and one can show that the branching ratio there is
\beq
B_c(\beta^*) = 1 - \frac{5}{4 \sqrt{7}} \approx 0.5275 .
\eeq
Note that as expected both model~I and model~II have $\gst=-\half$
away from the transition and $\gst^*=-\third$ at the critical point.

\begin{figure}[bth]
\caption[l]{Graph of the branching ratio, $B_c$, against the central
charge, $C$, for various models coupled to 2d gravity.}
\label{fig:Bcentral}
\begin{picture}(100,100)(0,0)
\centerline{\epsfbox{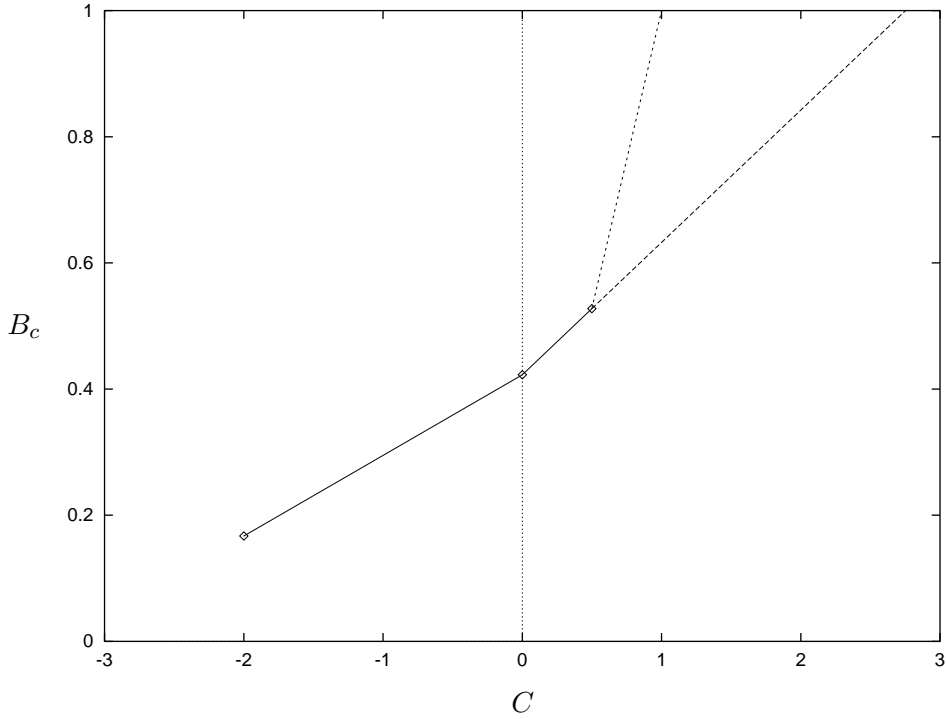}}
\end{picture}

\begin{picture}(10,10)(-10,-55)
$B_c$
\end{picture}

\begin{picture}(10,10)(-77,-15)
$C$
\end{picture}
\end{figure}

\section{Conclusion}
\label{sec:concl}
In this paper we have calculated the branching ratio for three
different models coupled to 2d gravity, in the form of rooted planar
\ph3 graphs. In figure~\ref{fig:Bcentral} are plotted the values of
$B_c$ against the central charge, $C$; for the case of the single spin
Ising model we used $B_c(\beta^*)$, since the graphs are most branched
at the phase transition.
We see that the graphs tend to become more
branched as the central charge increases. This is not unexpected since
many models coupled to gravity degenerate to branched polymers for
large enough central charge. One should perhaps note at
this point that $B_c$ is not a universal function of the central
charge, so that different models with $C\!=\!-2$ will in general have
different branching ratios. Thus the data point for the Gaussian model
is probably not very reliable when trying to determine the behaviour
of the multiple Ising model. For this reason we extrapolate using only the
last two data points and estimate that $B_c=1$ at $C=2.75$ (the
extrapolation is the right-most dotted line in figure~\ref{fig:Bcentral}). 
This is
likely to be an overestimate since the graph seems to be curving
upwards with $C$. Even if we ignore the data point for the Gaussian model
we know that for $C \to - \infty$, $B_c$ must tend
to a finite non-negative value and in this limit one might guess that
$B_c \to 0$; so that it seems likely that the graph will be curving
upwards in the region of interest. In our previous paper~\cite{paper4}
it was shown that regions of the phase diagram with $B_c=1$ correspond
to the branched polymer phase ($\gst=\half$) and its boundary
($\gst\ge 0$).
It is known that there is no branched polymer phase for
$C<1$ and thus the left-most dotted line indicates an approximate upper
limit on possible values of $B_c$. 
So we conclude that the curve of $B_c$ against $C$ probably lies
within the region bounded by the two dotted lines and hence that 
there exists a branched polymer phase
for central charges greater than some value $\overline{C}$ where $1
\le \overline{C} \ltsim 2.75$. Note that this estimate for $\overline{C}$ is
lower than previous cruder estimates~\cite{paper2,paper3}.
Our result is still consistent with the hypothesis $\overline{C}=1$,
although it is not possible to rule out the existence of an
intermediate region. Even though we are unable to distinguish between
these two possibilities it seems likely that any intermediate region
which exists would be quite narrow.
Further work will be required in order to determine more fully
the phase diagram for the multiple Ising model coupled to gravity and
to understand the nature of the $C=1$ barrier in general.

\subsection*{Acknowledgements}
The author would like to 
acknowledge the support of the Royal Society through
their European Science Exchange Programme.

\end{document}